\documentclass[12pt,epsf,psfig]{article}
\usepackage{amsmath,amssymb}
\usepackage{color,graphicx}

\begin{document}
\title{\bf  Anomalous magnetic and electric moments of $\tau$
    and lepton flavor mixing matrix in effective lagrangian approach}
\author{Jian-Qiang Zhang, Xing-Chang Song\\
    {\sl\small Institute of Theoretical Physics, Peking University, Beijing 100871, China}\\
    Wu-Jun Huo,  and Tai-Fu Feng\\
    {\sl\small Institute of High Energy Physics, Academia Sinica, P.O.Box 918-4,
    Beijing 100039, China}}
\date{}
\maketitle

\begin{abstract}
    In an effective lagrangian approach\,\cite{EM97} to new physics,
    the authors in ref.\,\cite{HL99} pushed tau anomalous magnetic
    and electric dipole moments (AMDM and EDM) down to $10^{-11}$
    and $10^{-25}~e~cm$ by using a Fritzsch-Xing lepton mass matrix
    ansatz. In this note, we find that, in this approach, there
    exists the  connection between $\tau$ AMDM and EDM  and the
    lepton flavor mixing matrix. By using the current neutrino
    oscillation experimental results, we investigate the parameter
    space of lepton mixing angles to $\tau$ AMDM and EDM. We can
    obtain the same or smaller bounds of $\delta\alpha_\tau$ and
    $d_\tau$ acquired in ref.\,\cite{HL99} and constrain $\theta_l$
    (the mixing angle obtained by long-baseline neutrino oscillation
    experiments) from $\tau$ AMDM and EDM.
\end{abstract}

\newpage
\section{Introduction}
    \label{S:intro}
    Some neutrino oscillation experiments in recent years indicate that
    neutrinos are massive and oscillate in flavor. That is to say,
    there exists lepton flavor mixing. Lepton flavor mixing is very
    important for solving some topics related to particle physics and
    cosmology, such as the lepton anomalous magnetic dipole moments
    (AMDM) and electric dipole moments (EDM) and lepton flavor
    violation (LFV) decays.

    Theoretically, the standard model predicts
    $\delta\alpha_{\tau}=1.1769(4)\times 10^{-3}$ and a very tiny
    $d_{\tau}$~\cite{CM98} from CP violation in the quark sector.
    Experimental analysis of the $e^{+}e^{-}\rightarrow \tau^+
    \tau^-\gamma$ process from L3 and OPAL collaborations gives that
    $\delta\alpha_{\tau}=0.004 \pm 0.027\pm 0.023$ and $d_{\tau}=
    (0.0 \pm 1.5 \pm 1.3) \times 10^{-16}
    ~e\,cm$\,\cite{T98}.
We also note that the authors in
ref. \cite{cal1,cal2} investigated  $g-2$ and raditive lepton decays with the
effective lagrangian approach. Moreover, in ref. \cite{HL99}, T. Huang, Z.H.
Lin and X.M. Zhang  firstly introduced the non-universal
effective operators and leaded the lepton flavor violation to the effective
approach.
 The authors of ref.\,\cite{EM97} presented
    an effective lagrangian to describe the effect of new physics
    and obtained the bounds of AMDM and EDM. By using a special
    lepton mass matrix ansatz\,\cite{FX96}, ref.\,\cite{HL99} pushed
    the anomalous magnetic and electric dipole moments of tau lepton
    down to
    \begin{equation} \label{E:TheoLim}
        \left|\,\delta\alpha_{\tau} \right| < 3.9 \times 10^{-11}, \quad \,\,\, \quad
        \left|\,d_{\tau} \right| < 2.2 \times 10^{-25}~e\,cm.
    \end{equation}
    However these results were obtained only in one Fritzsch-Xing ansatz
    which induces the nearly bi-maximal mixing pattern for atmospheric
    and solar neutrino oscillations.

    In fact, there are a lot of lepton mass matrices ansatz
    \cite{FX00} which are compatible with the current neutrino
    oscillation experimental data. Similar to the CKM matrix, the
    lepton mixing matrix can be also measured by experiments. However,
    at the present neutrino oscillation experiment level, it is
    difficult to put the precise values of lepton mixing matrix
    elements. From the several famous oscillation experiments, we can
    just obtain the range of these matrix elements. In the effective
    lagrangian approach, AMDM, EDM and lepton flavor violation are
    mainly related to the lepton mixing matrix. And thus we can find
    a connection between AMDM and EDM of tau lepton and the lepton
    flavor mixing matrix. As a result, either can we investigate the
    parameter space of lepton mixing angles from experimental data of
    $\tau$ AMDM and EDM, or can study the bounds of $\delta a_\tau$
    and $d_\tau$ imposed by the mixing angles obtained from the
    current neutrino oscillation experiments.

    In this note, we briefly talk about AMDM, EDM and LFV in the
    effective lagrangian approach introduced in Sec.\,2. In the sequent
    Sec.\,, the lepton mixing matrix and the numerically results are
    presented. And we summarize the subject in the last Sec.\,.

\section{Effective lagrangian with magnetic and electric dipole moments operators}
    \label{S:EffLag}

    To study magnetic and electric moments beyond Standard Model (SM),
    the authors in refs.\,\cite{EM97,HL99} considered an effective
    lagrangian approach to new physics:
    \begin{equation} \label{E:for3}
        \mathcal{L}_{Eff}=\mathcal{L}_{SM}+\frac{1}{\Lambda^{2}} \sum_{i}C_{i}\mathcal{O}_{i},
    \end{equation}
    where $\mathcal{L}_{SM}$ is the SM lagrangian, $\Lambda$ is the new
    physics scale, $\mathcal{O}_{i}$ are $SU_{C}(3) \times SU_{L}(2) \times
    U_{Y}(1)$ invariant operators and $C_{i}$ are constants which
    represent the coupling strengths of $\mathcal{O}_{i}$. A complete list
    of the operators can be found in Ref.~\cite{HY99}. Related to the
    anomalous magnetic moment of tau lepton, there are two
    dimension-six operators
    \begin{align}
        \mathcal{O}_{\tau B} &= \bar L_{\tau} \sigma^{\mu \nu} \tau_{R} \Phi
            B_{\mu \nu} \label{E:for4}\\
        \mathcal{O}_{\tau W} &= \bar L_{\tau} \sigma^{\mu \nu}
        \frac{\sigma^{i}}{2} \tau_{R} \Phi W_{\mu \nu}^{i} \label{E:for5}
    \end{align}
    where $\sigma^{i}$ is Pauli matrices (\,i=1,\,2,\,3\,), $L_{\tau}=(\nu_{\tau},\tau_{L})^T$
    the $\tau$ left-handed isodoublet, $\tau_{R}$ the right-handed singlet,
    $\Phi$ the Higgs scalar doublet, and $B_{\mu \nu}$ and $W_{\mu \nu}^{i}$
    are strengths of $U_{Y}(1)$ and $SU_{L}(2)$ gauge fields. Similarly,
    operators below are introduced to induce the electric dipole moments
    of $\tau$\,\cite{EM97},
    \begin{align}
        \tilde{\mathcal{O}}_{\tau B} &= \bar L_{\tau} \sigma^{\mu \nu} i \gamma_{5}
            \tau_{R} \Phi B_{\mu \nu} \label{E:for6}\\
        \tilde{\mathcal{O}}_{\tau W} &= \bar L_{\tau} \sigma^{\mu \nu} i \gamma_{5}
            \frac{\sigma^{i}}{2} \tau_{R} \Phi W_{\mu \nu}^{i}. \label{E:for7}
    \end{align}

    When $\Phi$ gets vacuum expectation value, operators $\mathcal{O}_{\tau
    B}$ and $\mathcal{O}_{\tau W}$ can give rise to the tau anomalous
    magnetic moment. After the broken of electroweak symmetry and
    diagonalization of the mass matrices of the leptons and the bosons,
    the effective neutral current couplings of the leptons to the photon
    $\gamma$ is
    \begin{equation} \label{E:for9}
        \mathcal{L}_{Eff}^{\gamma}=\mathcal{L}_{SM}^{\gamma}+ e
        \frac{1}{2 m_{\tau}}(-ik_{\nu}\sigma^{\mu \nu})S^{\gamma}
        \begin{pmatrix}
            \bar{e}\\
            \bar{\mu}\\
            \bar{\tau}
        \end{pmatrix}^{T}
        U_l^\dag
        \begin{pmatrix}
            0 &   &  \\
              & 0 &  \\
              &   & 1
        \end{pmatrix}
        U_l
        \begin{pmatrix}
            e\\
            \mu\\
            \tau
        \end{pmatrix},
    \end{equation}
    where
    \begin{equation} \label{E:for10}
        S^{\gamma}=\frac{2m_{\tau}}{e}
        \frac{\sqrt{2}v}{\Lambda^{2}} \left[ C_{\tau W}
        \frac{S_{W}}{2}-C_{\tau B} C_{W} \right],
    \end{equation}
    where $S_{W} \equiv \sin{\theta_{W}}$, $C_{W} \equiv \cos{\theta_{W}}$,
    and $\theta_{W}$ is Weinberg angle. Matrix $U_{l}$ is the unitary matrix
    which diagonalizes the mass matrix of the charged leptons. It is not
    measurable and dependent on the basis choice. For simplicity, we can
    choice the basis where neutrino is diagonal and work on neutrino mass
    basis. So, the unitary matrix $U_l^\dag$ here is just the lepton mixing
    matrix $V_l$,
    \begin{equation} \label{E:for11}
        V_{l}=U_{l}^{\dag}U_{\nu},
    \end{equation}
    where $U_{\nu}$ is the unitary transformation matrix that diagonalize
    the neutrino mass matrices $M_{\nu}$. By representing
    \begin{equation} \label{E:for12}
        \bar{V}=V_{l}
        \begin{pmatrix}
            0 &   &  \\
              & 0 &  \\
              &   & 1
        \end{pmatrix}
        V_{l}^{\dag},
    \end{equation}
    the decay width of $l \rightarrow l'+\gamma$ is given by
    \begin{equation} \label{E:for13}
        \Gamma(l \rightarrow l' \gamma)=\frac{m_{l}}{32 \pi}
        \left( \bar{V}_{ll'}e S^{\gamma}
        \frac{m_{l}}{m_{\tau}} \right)^{2},
    \end{equation}
    where $\bar{V}_{ll'}(l \neq l')$ are the non-diagonal elements
    of matrix $\bar{V}$ defined in Eq.~\eqref{E:for12}. The new
    physics contribution to the tau anomalous magnetic moment is
    given by~\cite{HL99}
    \begin{equation} \label{E:for14}
        |\,\delta \alpha_{\tau}|=|\,\bar{V}_{\tau \tau}S^{\gamma}|.
    \end{equation}
    A similar formula holds for the tau electric dipole moment. So we can find
    $\tau$ AMDM, EDM and LFV decays are all related to lepton mixing through
    the matrix ${\bar V}$.

\section{Lepton mixing matrix and  $\tau$  AMDM and EDM }
    \label{S:MixMat}
   The general form of the lepton flavor mixing matrix
    is~\cite{RG01}
    \begin{equation} \label{E:mix1}
        V_l=
        \begin{pmatrix}
            C_{12} C_{13} & -S_{12} C_{13} & S_{13}\\
            S_{13} S_{23} C_{12}+ C_{23} C_{12} e^{i\delta}
                & -S_{13} S_{23} S_{12}+ C_{23} C_{12} e^{i\delta}
                & -C_{13} S_{12}\\
            -S_{13} C_{23} C_{12} + S_{23} S_{12} e^{i\delta}
                & S_{13} C_{23} S_{12} + S_{23} C_{12} e^{i\delta}
                & C_{13} S_{23}
        \end{pmatrix},
    \end{equation}
    where $S_{ij} \equiv \sin{\theta_{ij}}$, $C_{ij} \equiv
    \cos{\theta_{ij}}$, $\theta_{ij}$ is the mixing angles between
    different flavors; and $\delta$ is CP phase.
    Putting aside the LSND experiments\,\cite{LSND}, the mixing angles measured
    by atmospheric\,\cite{Fu98}, solar\,\cite{SN96} and long-baseline\,\cite{CHOOZ}
    experiments actually correspond to the $\theta_{23}$, $\theta_{12}$,
    $\theta_{13}$ respectively. If neglecting the possible CP-violating phase, and
    substituting $\theta_{13}$, $\theta_{23}$, $\theta_{12}$ with $\theta_{l}$,
    $\theta_{a}$, $\theta_{s}$, we can present the lepton mixing matrix
    as~\cite{FX00,BP98}
    \begin{equation} \label{E:mix2}
        V_l=
        \begin{pmatrix}
            C_{l} C_{s} & -S_{s} C_{l} & S_{l}\\
            S_{l} S_{a} C_{s}+ C_{a} C_{s}
                & -S_{l} S_{a} S_{s}+ C_{a} C_{s} & -S_{s} C_{l}\\
            -S_{l} C_{a} C_{s} + S_{a} S_{s}
                & S_{l} S_{s} C_{a} + S_{a} C_{s} & S_{a} C_{l}
        \end{pmatrix},
    \end{equation}
    where $S_{a} \equiv \sin{\theta_{a}}$, $C_{a} \equiv
    \cos{\theta_{a}}$, and so on.
    The present experimental data favor $\sin^{2}{2
    \theta_{a}}>0.8$ or $\theta_{a} \sim 32^{\circ}-45^{\circ}$,
    $\sin^{2}{2 \theta_{l}}<0.1$ or $\theta_{l} \sim 0^{\circ}-
    9.2^{\circ}$. There are two different possible cases to the
    solar neutrino oscillation: the long wave-length vacuum
    oscillation with $\sin^{2}{2 \theta_{s}} \approx 1$,
    or $\theta_{s} \sim 45^{\circ}$; and the matter-enhanced
    oscillation (MSW mechanism~\cite{MS85}) with
    $ \sin^{2}{2 \theta_{s}} \sim 10^{-3}-10^{-2}$ (small-angle
    solution) and $\theta_{s} \sim 1^{\circ}-3^{\circ}$ or with
    $ \sin^{2}{2 \theta_{s}} \sim 0.65-1$ (large-angle
    solution) and $\theta_{s} \sim 27^{\circ}-45^{\circ}$.

    From Eqs.\,\eqref{E:for12}-\eqref{E:for14} and \eqref{E:mix2},
    we get the explicit form of the anomalous magnetic moment
    expressed in $\theta_{l}$, $\theta_{a}$ and $\theta_{s}$,
    \begin{equation} \label{E:mix3}
       |\,\delta \alpha_{\tau}|=
            \frac{m_{\tau}}{e m_{\mu}}
            \left| \frac{C_{l} S_{a}^{2}}{S_{l} S_{s}} \right|
            \sqrt{\frac{32 \pi \Gamma_{(\mu \rightarrow e
            \gamma)}}{m_{\mu}}}~.
    \end{equation}
    By using  the current experimental upper limits,
    $BR (\mu^{-} \rightarrow e^{-} \gamma )< 4.9 \times 10^{-11}$\,\cite{PDG98},
    we investigate the relation between $\delta a_\tau$ and $d_\tau$
    and the mixing angle $\theta_l$ which are shown on Figs. 1, 2, 3 and 4.
    In the numerical calculation, we take $\theta_a =32^{\circ}$ and
    $45^{\circ}$ respectively, as well as $\theta_s =1^{\circ}$,
    $27^{\circ}$, $45^{\circ}$.

    In the Figs, such as Fig.\,1,  all three curves, which correspond
    to $\theta_s =1^{\circ}, 27^{\circ}$ and $45^{\circ}$
    respectively, as well  as $\theta_a =32^{\circ}$, are all decrease
    to the lepton mixing angle which constrained by the long-baseline
    neutrino oscillation experiments. On the one hand, Considering the
    bound of $\delta a_\tau$ in Eq.\,\eqref{E:TheoLim}, we find that
    the curve A can
    be excluded. That is to saw, $\theta_s$ will be not in the region
    $1^{\circ} -3^{\circ}$.  This result is suited to large mixing
    angle MSW solution ($\theta_s \sim 27^{\circ} -45^{\circ}$) which
    is favored by the recent data of the Super-K and SNO. For curve B
    (or C), the regions in which $\theta_l < 1.8 ^{\circ}$ (or
    $\theta_l < 3 ^{\circ}$) can be excluded. The similar analysis can
    be used to the other three Figs. On the other hand, from the
    lepton flavor mixing angles constrained by the experiments
    directly, we can obtain the bounds on $\delta a_\tau$ and $d_\tau$
    same as or smaller than these from the ansatz in Eq.\,\eqref{E:TheoLim},
    if $\theta_l$ lies in the scope about from $1.8^{\circ}$ to
    $ 9.2^{\circ}$.  In other words, by using the effective lagrangian
    approach and the lepton mixing matrix, we can push $\tau$ AMDM and
    EDM down to very tiny values.

\section{Summary}

    We extend the work in ref.\,\cite{HL99} to bound the AMDM and
    EDM of $\tau$ in the effective lagrangian by considering the
    current experimental lepton mixing data. The experimental
    limits on $\mu\to e\gamma$ can put strong limits on $\tau$
    AMDM and EDM same as obtained in ref.\,\cite{HL99}. Our results
    are compatible with the large mixing angle MSW solution  which is
    favored by the recent data of the Super-K and SNO. The AMDM and
    EDM of $\tau$ can also give some constraints on the lepton mixing
    angles $\theta_s,\theta_a$ and $\theta_l$ by solar, atmospheric
    and long-baseline  neutrino experiment respectively.

\section*{Acknowledgments}
    This work is supported in part by National Natural Science
    Foundation of China and Doctoral Programme Foundation of
    Institute of High Education of China. One of the authors(W.J.H)
    acknowledges supports from the Chinese Postdoctoral Science
    Foundation and CAS K.C. Wong Postdoctoral Research Award Fund.
    We grateful to Prof. T. Huang, X.M Zhang and Z.Z Xing for useful
    discussions.

\begin{figure}[ht]
    \begin{minipage}[t]{1\textwidth}
        \scalebox{1}[1]{\includegraphics*[50pt,305pt][370pt,550pt]{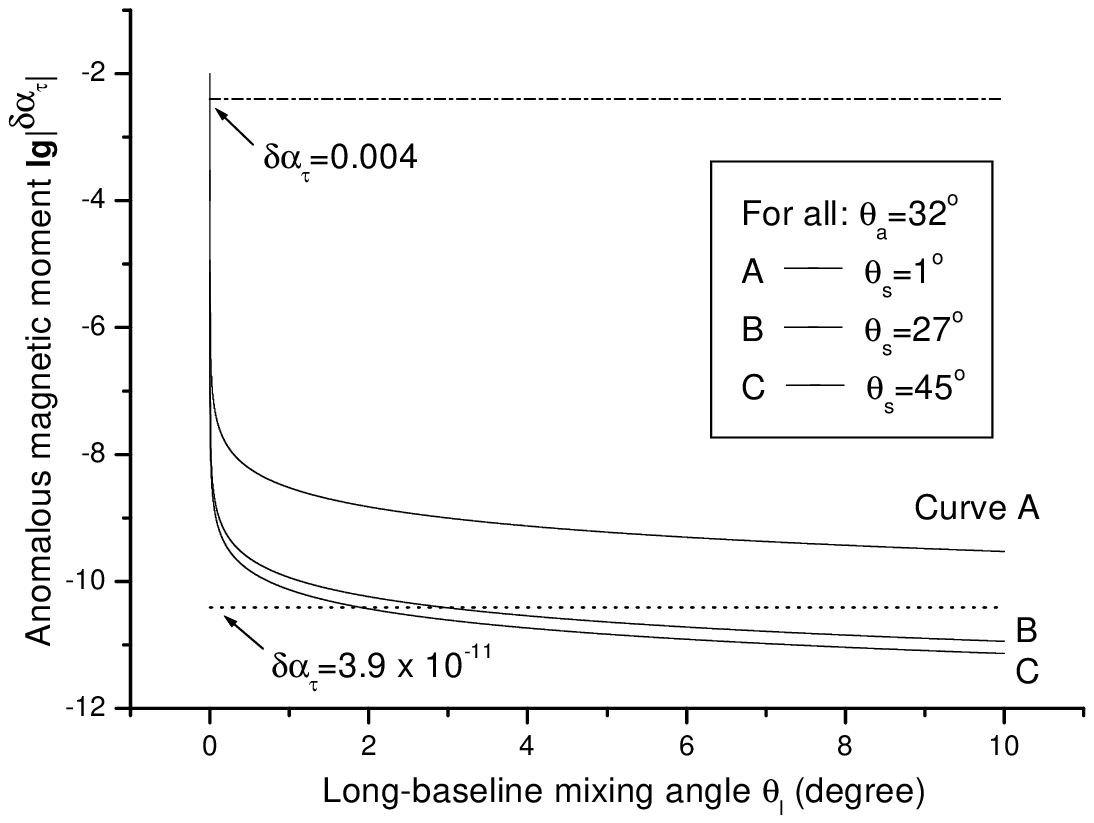}}
        \caption{Fig of $|\delta\alpha_\tau |$ to $\theta_l$ when $\theta_a
            =32^{\circ}$ , as well as $\theta_s =1^{\circ}$, $27^{\circ}$,
            $45^{\circ}$ respectively.}
        \label{Fig:fig1}
    \end{minipage}
\end{figure}
\begin{figure}[ht]
    \begin{minipage}[t]{1\textwidth}
        \scalebox{1}[1]{\includegraphics*[50pt,305pt][370pt,550pt]{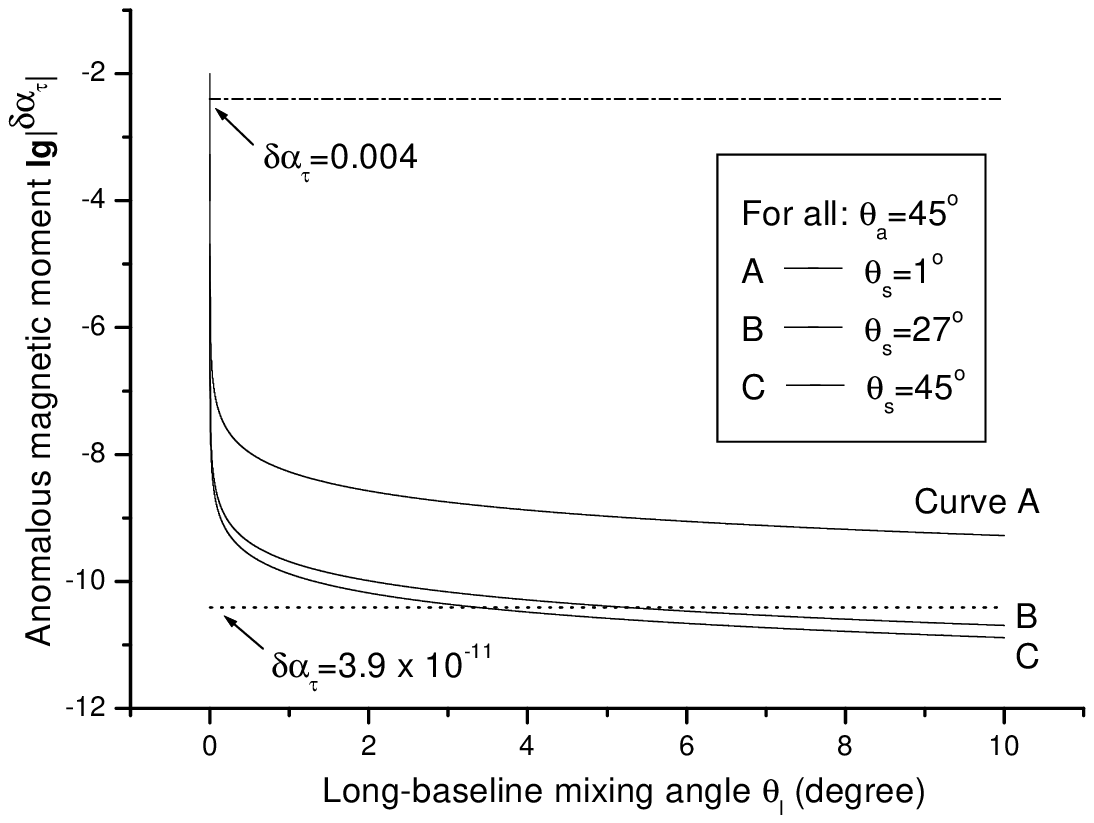}}
        \caption{Same as Fig. 1 but $\theta_a =45^{\circ}$.}
        \label{Fig:fig2}
    \end{minipage}
\end{figure}
\begin{figure}[ht]
    \begin{minipage}[t]{1\textwidth}
        \scalebox{1}[1]{\includegraphics*[50pt,305pt][370pt,550pt]{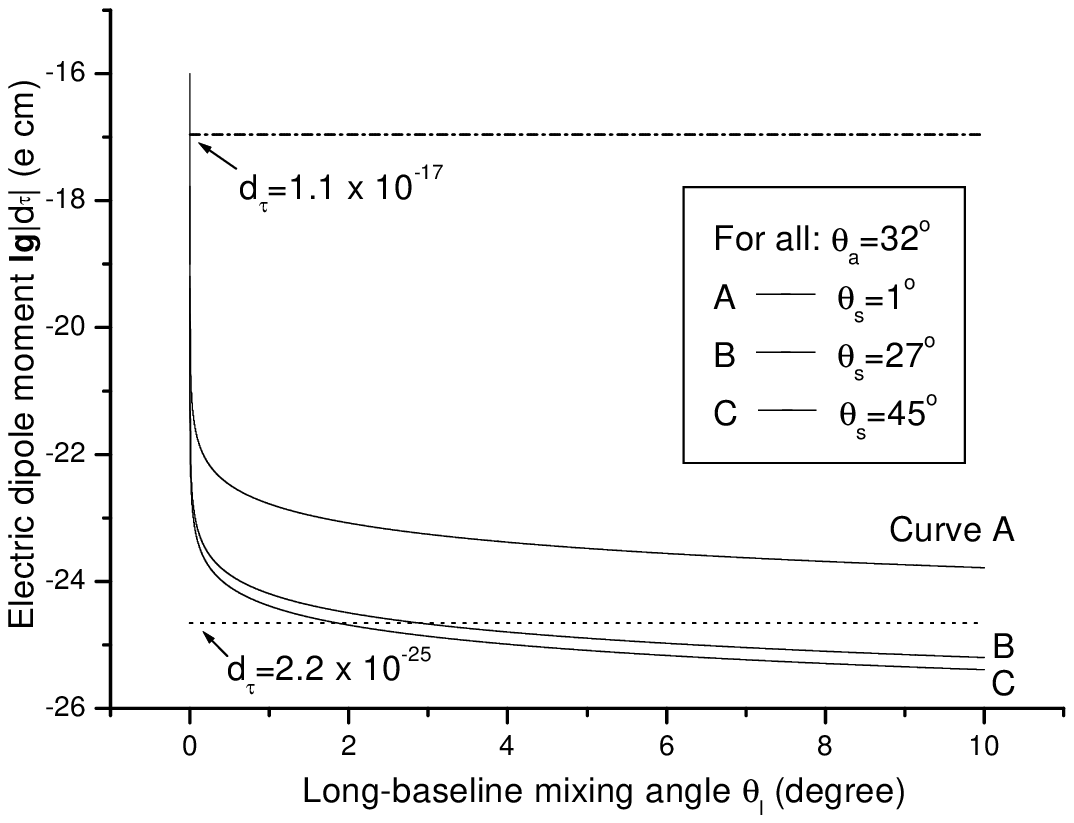}}
        \caption{Fig of $|d_\tau |$ to $\theta_l$ when $\theta_a =32^{\circ}$,
            as well as $\theta_s =1^{\circ},\,\,27^{\circ},\,\,45^{\circ}$
            respectively.}
        \label{Fig:fig3}
    \end{minipage}
\end{figure}
\begin{figure}[ht]
    \begin{minipage}[t]{1\textwidth}
        \scalebox{1}[1]{\includegraphics*[50pt,305pt][370pt,550pt]{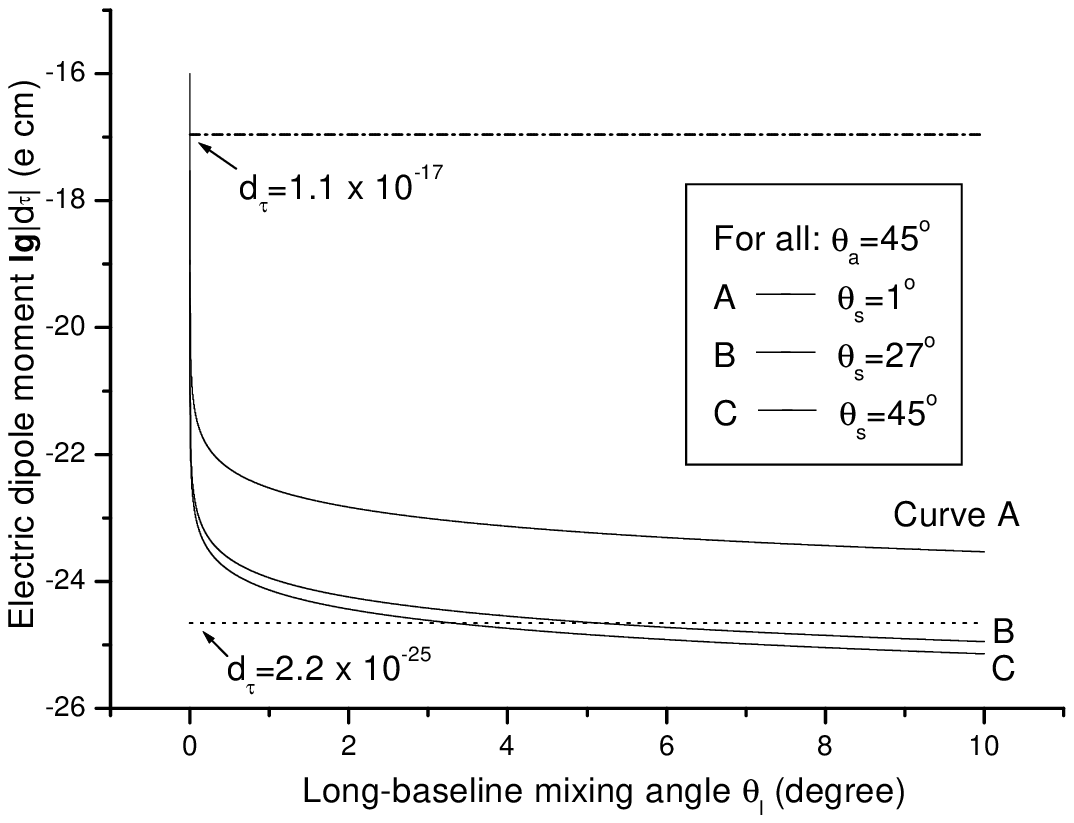}}
        \par\vspace{0pt}
        \caption{Same as Fig. 3 but $\theta_a =45^{\circ}$.}
        \label{Fig:fig4}
    \end{minipage}
\end{figure}

\end{document}